# Relationships between Journal Publication, Citation, and Usage Metrics within a Carnegie R1 University Collection: A Correlation Analysis

William H. Mischo, Mary C. Schlembach, Elisandro Cabada[1]

This study examines the correlational relationships between local journal authorship, local and external citation counts, full-text downloads, link-resolver clicks, and four global journal impact factor indices within an all-disciplines journal collection of 12,200 titles and six subject subsets at the University of Illinois at Urbana-Champaign (UIUC) Library. While earlier investigations of the relationships between usage (downloads) and citation metrics have been inconclusive, this study shows strong correlations in the all-disciplines set and most subject subsets. The normalized Eigenfactor was the only global impact factor index that correlated highly with local journal metrics. Some of the identified disciplinary variances among the six subject subsets may be explained by the journal publication aspirations of UIUC researchers. The correlations between authorship and local citations in the six specific subject subsets closely match national department or program rankings.

## Introduction

There has been a great deal of interest in attempting to determine correlational relationships between various individual journal title metrics within a collection, including local publication, citation, and usage (now download) measures and journal global impact factor measures. Gathering the raw measurement numbers associated with specific journals and establishing relationships between these variables can serve to inform a library's collection development and management decisions, including journal subscription, cancellation, and retention decisions. From a public service and subject liaison perspective, this data can be used to construct a knowledgebase identifying departmental and faculty research concentrations and areas of focus.

The data can also be used to indicate if the collection is adequately supporting the research and instructional needs of faculty and students. It can be used in the generation of a library's core journal list, which can provide an evidence-based listing of journals necessary to meet the instructional and research needs of the library's primary constituents. [1]

---

[1] *William H. Mischo is Berthold Family Professor Emeritus in Information Access and Discovery and Head, Grainger Engineering Library Information Center, University of Illinois at Urbana-Champaign; Mary C. Schlembach is Chemistry & Physical Sciences Librarian, Head, Chemistry Library, University of Illinois at Urbana-Champaign; Elisandro Cabada is Medical & Bioengineering Librarian, Grainger Engineering Library Information Center, University of Illinois at Urbana Champaign; emails: w-mischo@illinois.edu; schlemba@illinois.edu; cabada@illinois.edu ;* 

The raw publication, citation, and usage data and correlation measures can assist in developing mechanisms to calculate a journal's overall local composite value and can contribute to providing more data-driven assessments of a library's journal collections. Collecting and correlating authorship, citation, and usage data can also allow patterns of journal use to emerge, resulting in a more accurate picture of journal value than cost-per-use calculations or other value gathering methods. Data gathered in this process can also be used to defend any local administrative tax that could be applied to academic departments or colleges to fund the library.

Libraries are also interested in determining the degree to which any one of the local journal metrics, particularly full-text downloads or local citations, can be used as a proxy for predicting any of the other values. This might allow, for example, predictive statements about publication numbers or citations to be made from usage numbers, or vice versa, and for one measure to serve as a predictive proxy for another measure. If this were uniformly the case, libraries could focus on collecting one or two types of measurements and be certain that the other local metrics would be proportionate.

In the same way, libraries are also interested in determining to what extent the journal global impact factor indices, such as the Institute for Scientific Information (ISI) Journal Citation Reports (JCR) Journal Impact Factor (JIF), (abbreviated as ISI JCR) correlate with local citation, publication, and usage metrics. A high correlation would in theory allow impact factor index values to be used in journal collection decisions or serve as a proxy for the local data variables.

Libraries have local usage data available in the form of the *Counting Online Usage of Networked Electronic Resources* (COUNTER) full-text download usage reports. [2] COUNTER full-text download data is provided by commercial and professional society publishers in the form of spreadsheets for specific journals, giving monthly and yearly download data. Local publication and citation information is commonly available via several tools, among them the *Local Journal Utilization Report* (LJUR) from ISI, Scopus API extracted data, the SciVal PURE current researcher profile information system, the Symplectics Elements platform, and other research management resources.

This paper examines the relationships between specific journal title publication, citation, usage, and impact metrics from 17,934 journals in all disciplines, including 12,200 active titles, gathered from scholarly activities involving researchers at the University of Illinois at Urbana-Champaign, a Carnegie R1 university. The study analyzes and calculates the correlations over five years of local journal authorship numbers; five years of local and external citation counts; two years of full-text downloads, two years of link-resolver data, and the values from four global journal impact factor indices. It also examines correlation pair values for the journals in six subject subset areas: engineering; life sciences; social sciences; chemistry; history and philosophy; and literature. Several of these are monographic focused literatures and were included to present a more comprehensive scholarly communication model.

## Literature Review

There is a long and rich literature on journal publication, citation, and usage metrics, particularly in the area of citation analysis which is defined as the examination of citations from journal articles, dissertations, or other publications to determine trends and patterns of use. [3] Ashman reviewed and categorized 88 studies on citation analysis culled from the library literature published from 1995 to 2008, categorizing the articles into nine types of literature profiles in the areas of public service, assessment, and collection-related areas. [4] Hoffmann and Doucette reviewed 34 articles on citation analysis methodologies published from 2005 to 2010 and found that the articles typically did not provide enough information to make them adaptable for practical collection management decisions. [5]

Many studies have examined the relationships between citation and usage, in both local and global settings, first with print collections, then later with e-journal collections. Research relating to this topic appears in both the library and informetrics literature.

Surveying the studies on the relationship between citations and usage, McGillivray and Astell note that "these (studies) have not produced a definitive answer." [6] Pastva et al examined the literature on citation and usage analysis and stated that "some studies found a significant correlation between citation and usage data, while others found no significant correlation, highlighting the importance of methodology and local citation behaviors." [7]

Several studies have looked at the relationship between usage and citation at the article or paper level and sometimes at both the article and journal title levels. Brody, Harnad, and Carr examined download and citation patterns at the paper level within the ArXiv.org e-print archive and found that the number of times an article was read was related both to the number of times it was cited and the age of the article. The authors also determined that short-term Web usage predicted medium-term citation impact. [8] Kurtz et al examined citation rates and readership rates with respect to publication date within the NASA Astrophysics Data System and developed a model for the relationship between reads and cites which incorporates obsolescence and derives a citation function that is based on several components of a usage function. [9] In a review of usage and citation studies, Kurtz and Bollen assert that the relationship between usage and citation is complex and state: "with the accurate description of use being so complex, it is perhaps not surprising that the relation between use and citation has not been convincingly established." [10] They describe the difficulty in comparing usage information at the article level with citation histories and show that the interpretation of usage frequency as a function of publication date is quite complex. [11] Schloegl and Gorraiz found that there were differences between downloads and citations in terms of obsolescence characteristics, where the half-life of the articles that are downloaded and the median cited half-life are significantly different. They found that the average cited half-life was 5.6 years and the mean usage half-life was 1.7 years, complicating the correlation relationship. [12]

Many of the studies examining the relationships between citation, usage, publication, and impact factor metrics have been carried out in specific subject disciplines and have typically covered a small subset of a discipline's journals. [13] In addition, several studies have found that the correlation between citation and usage data is dependent on subject discipline. [14]

Several previous studies have examined broader correlations between publication, usage, citation, and impact factor metrics within a library environment. Duy and Vaughan found a significant correlation between electronic journal usage and both LJUR local citation counts and library shelving counts for 112 chemistry and biochemistry journals but found no significant correlation between the ISI JCR impact factor data and local electronic journal usage. [15] McDonald, using subsets from 1,521 journal titles from the California Institute of Technology Library, found that print journal usage and, later, online journal usage was a valid predictor of local citation rates in journals. [16]

De Groote, Blecic, and Martin examined 2,619 health science journals and found high correlation values between download data, link-resolver data, and local citation rates. [17] They concluded that link-resolver data were a good predictor of usage statistics in this environment. However, Gallagher, Bauer, and Dollar found that the usage data captured by link resolvers represented less than 10% of the total e-journal usage as identified by vendor download data. [18]

Chew et al collected metrics data from 700 e-journals within 12 disciplinary subject areas at the University of Minnesota and analyzed correlation values. [19] The study found marked disciplinary variation in the resulting correlations and also significant discrepancies between Scopus and ISI Web of Science calculated values. Some of the sample sizes were quite small.

Pastva et al conducted a citation and usage analysis over 33,000 articles (from an indeterminate number of journal titles) published between 2007 and 2016 by researchers in the Feinberg College of Medicine at Northwestern University. [20] The correlations they derived between journal title usage and citing data were fairly weak and open to interpretation.

In a study of the University of Houston School of Communication faculty publications, Gao analyzed correlations among citation count, journal impact factor values, and journal usage. [21] The journal sample sizes for the studied factors ranged from 147 journal titles to 108 titles. Gao found significant correlations between journal impact factor values and journal usage but no correlation between citation count and impact factor.

Several studies have been performed in research or vendor settings. In a comprehensive scientific impact analysis, Bollen et al studied 7,675 journals and compared a number of journal citation and usage measures derived from usage log data from the 2008 Los Alamos Metrics from Scholarly Uses of Resources (MESUR) Project with several external impact factor measures. [22] The authors performed a principal component analysis over a 39x39 factor correlation matrix and found 10 usage-based measures that appear to be stronger indicators of scientific prestige than the ISI JCR and other citation-based impact factor measurement systems. Bollen et al comments that: "these results should give pause to those who consider the JIF (ISI JCR) the gold standard of scientific impact." [23] An earlier study by Bollen et al (2005) also questioned the validity of the ISI JCR as a valid assessment of journal impact and suggested that usage-based measures were more accurate on a local level. [24]

Gorraiz, Gumpenberger, and Schloegl looked at the use of citation and download global data from 362 ScienceDirect journals over 10 years covering four subject disciplines: arts and humanities, computer science, economics and finance, and oncology. [25] They found that the disciplines with the highest citation rates are not those with the highest download rates and the proportion of downloaded documents is dramatically higher than the proportion of cited documents. The authors claim that citations are often insufficient to assess the impact of the research output in many disciplines and downloads do not necessarily measure actual usage but must be considered as a complement to the "bibliometric citation-restricted horizon". [26]

Elkins et al examined the correlation between four journal impact factor indices, including the ISI JCR JIF, Eigenfactor's article influence index, SCImago's journal rank index, and Scopus' trend line index. [27] Paired values for the four all showed strong correlations between the four impact factor indexes.

Moed and Halevi carried out a detailed analysis of the relationship between downloads and citations by examining 62 journals from the Elsevier ScienceDirect repository, finding large differences in the degree of correlation between downloads and citations across various subject fields. [28] They examined the correlations at both the journal and article levels finding that downloads were a good predictor of citations but that citations were a less valid predictor of downloads.

In addition, tools have been developed for library managers that are designed to aid in evaluating journal collections using journal title metrics. The California Digital Library developed the Weighted Value Algorithm (CDL-WVA) dashboard for collection selectors. Knowlton, Sales, and Merriman found that faculty selection differed significantly with the bibliometric values provided by the CDL-WVA tool. [29] The Canadian Research Knowledge Network (CRKN) also utilized the CDL-WVA assessment tool in examining consortial packages of publisher journals. [30]

## Methodology

The UIUC is a Carnegie R1 university with over 34,000 undergraduates, over 17,000 graduate students, and 1,900 tenure-system faculty, offering degrees in over 150 programs. In 2019, UIUC awarded almost 14,000 degrees, including 874 PhDs. The UIUC Library supports this wide variety of instructional and research programs with comprehensive journal subscriptions from all major commercial and professional society publishers. In 2021, the UIUC Library supported 108 subject or central collection funds on a $19.5 million materials budget.

The goal of this study was to examine the relationships between a number of local journal title publication, citation, link resolver clickthroughs, and usage measures within a large research university setting and calculate the correlations of these local metrics with four global impact factor indices. Examining these local journal metrics along with the global impact factor data assists the library to better determine the scholarly activities of UIUC researchers and to accurately characterize journal value measures for collection development and retention purposes. To obtain the local publication and citation data for this analysis and correlation of journal research activity metrics, the University Library purchased the 2017 Local Journal Utilization Report (LJUR) for UIUC from ISI, now owned by Clarivate.

The LJUR data provides summary information on UIUC researcher journal title authorship and citation numbers for the database of journals covered by the ISI platform. The LJUR data covers all academic disciplines and provides local publication and citation data at the journal title level for the journals covered by the extended ISI source list that is comprised of the journals covered by the former Science Citation Index, the Social Sciences Citation Index, and the Arts and Humanities Citation Index. The LJUR database covers the years 1981 through 2017 and includes UIUC publication and citation data from 17,934 journals.

The LJUR data is packaged as a Microsoft Access relational database containing five tables: (1) a journal list of 17,934 titles with columns for standard title abbreviation, ISSN, active or inactive status, and publisher; (2) a source publication table of 8,587 journal titles that UIUC authors have published in from 1981 through 2017 with columns for the total number of articles published and published articles for individual years from 1981 through 2017; (3) several tables of UIUC author local citation numbers for 15,785 journals with total times cited and citations by year; (4) a table of 21,423 journal titles (with 14,140 unique titles) including a title and an ISI subject descriptor; and (5) several tables of 14,338 journal titles showing the number of times outside authors have cited articles written by UIUC authors.

For this study, the data from the LJUR was used as a base to construct a journal title master table within a relational database that contained the 17,934 LJUR journal titles list, including the ISSN and EISSN numbers, the publisher information, and the active/inactive designation. Several scripts were written that extracted publication and citation information from the other LJUR database tables and added this data as additional columns into the master table. The columns that were added contained the total number of locally authored articles in each journal for a five-year period from 2013 to 2017, the total number of local citations from UIUC authors for each title from 2013 to 2017, and the total of external citations to UIUC authored articles for each journal from 2013 to 2017. There were 12,200 journals in the master table that the LJUR designated as active titles.

In order to better process and add the data for the additional usage, impact factor, and SFX (the Ex Libris local link resolver used by libraries) clickthrough numbers to the table of 17,934 journal records, additional ISSN and EISSN numbers for individual titles were added as columns using data from the Australian Research Council Excellence in Research for Australia (ERA) 2018 Journal List and ISSN/EISSN data from the Scopus source journal list. This provided a more comprehensive list of ISSNs and EISSNs that could be used as linking keys for matching journal titles and extracting download data from COUNTER tables and the impact factor data from the various services.

Subject descriptors were added as a column to the journal title master table to more accurately identify and extract subsets of journal titles by subject disciplines. A script was written to extract the ISI subject descriptors from the appropriate LJUR table which contained subject terms drawn from five research areas described by 251 subject categories in the ISI Web of Science descriptor scheme. To augment these LJUR Web of Science descriptors, subject terms from Scopus were added as a separate column to the journal title master table. Scopus journal titles are classified under four broad subject clusters which are further divided into 27 major subject areas and 300+ minor subject areas. The Scopus subjects were added using ISSN and EISSN numbers as the linking key. The two subject columns were used in SQL statements on the journal title master table to retrieve relevant journals in six disciplinary categories covering engineering, chemistry, social sciences, biosciences, history/philosophy, and literature subsets.

In addition, within the master title list a column was added for SFX link resolver requests and clickthroughs for the years 2017 and 2018.

Four global impact indices were also used in the analysis. All the impact factor indices use global citation statistics to assign a value to individual journal titles typically calculated by taking the number of cited articles in a journal over a specific time period and dividing that number by the number of articles published in that same time period. The impact factor indices used in this analysis were: the ISI JCR JIF five-year impact factors from 2018; the SCIMago Journal Rank (SJR) values which use average citations within a subject field, the quality of citing journal, and a page rank algorithm on top of the usual measurement of citations divided by articles; the Eigenfactor scores from 2018 in which citations from highly ranked journals are weighted to generate a higher citation score than those from poorly ranked journals and normalized the journal scores by rescaling the total number of journals in the ISI JCR; and the Elsevier 2018 SNIP (Source Normalized Impact per Paper) values, which weights citations based on the total number of citations in a subject field over three years. All of these additional impact factor values have been added as separate columns in the journal title master entries using ISSN and EISSN numbers as the linking key.

For the usage data, the study utilized the publisher-provided COUNTER full-text download usage reports from 35 commercial and professional publishers and four aggregators – EBSCO, ProQuest, Ovid, and JSTOR. The publisher list includes all the major commercial publishers, e.g., Elsevier, Wiley, Springer-Nature, and Taylor and Francis, and many professional society publishers. The COUNTER data, in the form of spreadsheets of specific journal monthly and yearly full-text download data, was uploaded as individual tables in a separate companion publisher information relational database. Each publisher table contained COUNTER JR1 full-text usage report statistics from either 2015 or 2018. Over 44,000 journal titles are represented in the 39 COUNTER supplied publisher tables in the 2015 data and over 45,000 titles were in the 2018 data with 31,918 unique journal titles represented over both years. The duplicates are often titles appearing in both the publisher and aggregator tables. From there, scripting programs were written to move the 2015 and 2018 COUNTER downloaded data into an aggregated column for each specific matching journal in the journal title database master table. If a journal title appeared in more than one COUNTER table (for example a publisher and an aggregator), the numbers were added together to obtain a total number of downloads for that journal title. The journal titles were matched in the journal title master table using ISSN and EISSN numbers as linking keys.

There were 10,604 of the 17,934 journals in the LJUR all-discipline corpus that had COUNTER download numbers available and 9,190 of the 12,300 active journal titles with available download data, so the COUNTER statistics covered a high majority of the journal titles in this study.

All the raw data used in this analysis, in the form of relational database tables with multiple columns, is being made available in the UIUC Library's Illinois Data Bank dataset repository under https://doi.org/10.13012/B2IDB-6810203_V1. In addition, the processing scripts and Pearson correlation code is available at https://doi.org/10.13012/B2IDB-0931140_V1.

## Processing

The correlation processing was set up to examine nine journal metric indicators: the LJUR supplied local authorship, LJUR local citation, LJUR external citation, full-text COUNTER download usage, link resolver clickthrough results, and four global journal impact service indexes. The numeric values for these nine indicators are all stored as columns in the records in the journal title master table. Note that this correlation analysis was carried out at the journal title level and does not include any analysis at the journal article or paper level.

A web interface over the database master journal title table was created with a search function that allowed retrieval by journal title, publisher, and subject and sorting capabilities by each of the journal title metrics. The web site tool serves an administrative and search function. It displays data records on single or groups of journal titles with the journal title metrics and can inform subscription, retention, and cancellation decisions, assist liaison librarians in understanding department and group research fronts, and contribute to the identification of core journal lists.

There were 12,200 designated active journal titles in the list of 17,934 titles from the LJUR database at the time of the analysis. The correlations over the metric data elements were carried out on both the 12,200 active titles and the entire corpus of 17,934 journal titles. There was essentially no difference in the correlation values of the active titles analysis and the values on the entire corpus. Because the study was using a two-year total of download data and five-year totals of local authorship and local citation, it was determined that the 12,200 active journal set would serve as a more accurate base sample for correlation calculations. The correlation analyses were carried out over the complete all-discipline set of 12,200 journal titles and over six subsets comprised of engineering journals; chemistry journals; social science journals; biosciences journals; history and philosophy journals; and literature journal titles.

A server-side correlation generator script was written that produced a web site dashboard that allowed the authors to select the desired journal title metric indicators from the nine options and select either the all-disciplines journal corpus or one of the six disciplinary subset areas. The correlation generator produces a series of two-way correlations on the selected journal title indicators, producing a maximum of 26 (nine values taken two indicators at a time) pairwise correlation values – or fewer if less of the nine indicator factors are chosen. The correlations can be run over the 17,934 journal title corpus and also the 12,200 active journal title subsets.

The correlation generator calculates Pearson's R values for the two pairwise data points. Pearson's R gives values from -1 to 1 where -1 is a perfect negative correlation; 0 is no correlation; and 1 is a perfect positive correlation. Pearson's is intended to be used in situations where the raw numeric data is available, as in this case. Several previous studies used the Spearman's rho correlation statistic in cases where ranked data, not numeric data, was available. All the Pearson's values in this analysis are significant at the $p < .001$ or lower value.

## Global Impact Factor Measures

Table 1 shows the pairwise correlation analysis over the journal title values from the four global impact factor indices, using the 12,200 LJUR 2017 active journal titles as the base. Three of the global impact factor indices journal title values showed a high correlation to each other: the *ISI JCR* and *SNIP* (N=8121, R= 0.7674); the *ISI JCR* and *SJR* (N=8136, R=0.877); and the *SNIP* and *SJR* (N=11669, R=0.7438). While the SNIP, SJR, and the ISI JCR correlate highly with each other, the normalized Eigenfactor stands out as not correlating highly with any of the other three impact factor indices: *ISI JCR* and *Eigenfactor* (N=8348, R=0.4346); *SNIP* and *Eigenfactor* (N=8127, R=0.2392); *SJR* and *Eigenfactor* (N=8142, R=0.398).

| Table 1: Correlations Between Impact Factor Measures | | | |
|---|---|---|---|
| Impact Factor | SJR | ISI JCR | Eigenfactor |
| SNIP | N=11669 R=0.7438 | N=8121 R=0.7674 | N=8127 R=0.2392 |
| SJR | | N=8136 R=0.877 | N=8142 R=0.398 |
| ISI JCR | | | N=8384 R=0.426 |

Table 2 presents the correlations between local publications, citations, and downloads with the three highly correlated global impact factor indices: ISI JCR, SNIP, and SJR. None of the three impact factor indices exhibited a significant R value with the publication, citation, or usage measures for the cohort of journals in the study. This reinforces results obtained in numerous studies that show that the ISI JCR is often not a useful measure for local citation and publication activities and typically cannot serve as a proxy for local scholarly communication measures. [31]

| Table 2: Correlations Between Impact Factors and Local Publication, Citation and Usage | | | |
|---|---|---|---|
| Impact Factor | Publication | Local Citations | Downloads |
| SNIP | N=11676 R=0.0849 | N=11676 R=0.1397 | N=8935 R=0.1915 |
| SJR | N=11721 R=0.134 | N=11721 R=0.245 | N=8973 R=0.318 |
| ISI JCR | N=8384 R=0.1122 | N=8384 R=0.2364 | N=6171 R=0.3446 |

Based on these results, only the normalized Eigenfactor Score values were used in the remaining analysis, along with the five local publication, citation, link resolver, and usage indicators. Chew et al found that the Eigenfactor and SNIP (but not the ISI JCR) values provided significant correlations in certain disciplines with local publication and citation data. [32] They found local authorship and impact factor values correlated strongest with Eigenfactor but also with SNIP in several disciplines.

## Correlations Over the Journal Title Metrics

Table 3 shows the correlation results from the 12,200 active journal titles and journal value indicators for the all-disciplines analysis. Of the 15 pairwise value combinations (six items taken two at a time), only three Pearson R values are below .5 (shown in red text): the pair *SFX clickthroughs* and *outside citations* (N=11709, R=0.4351); *SFX clickthroughs* and *articles written* (N=11709, R=0.4941); and *downloads* and *outside citations* (N=9190, R=0.4959). All other R values are above .5 (shown in blue text) with the next lowest being the pair *downloads* and *articles written* (N=9190, R=0.5282), *SFX clickthroughs* and *locally cited* (N=11709, R=0.5863), and *normalized Eigenfactor* and *articles written* (N=8408, R=0.5937). All the other values are R=.64 or higher.

| Table 3 Correlation Results from 12,200 active journal titles in all Disciplines and Subjects | | | | | |
|---|---|---|---|---|---|
| | UIUC Author Cited by outside authors | Articles written by UIUC authors | Downloads of articles by UIUC users | SFX Clickthroughs | Normalized Eigenfactor |
| Locally Cited by UIUC authors | N= 12200 R=0.7613 | N= 12200 R=0.7698 | N= 9190 R=0.7843 | N= 11709 R=0.5863 | N= 8408 R=0.7858 |
| UIUC Author Cited by outside authors | | N= 12200 R=0.7907 | N= 9190 R=0.4959 | N= 11709 R=0.4351 | N= 8408 R=0.6429 |
| Articles written by UIUC authors | | | N= 9190 R=0.5282 | N= 11709 R=0.4941 | N= 8408 R=0.5937 |
| Downloads of articles by UIUC users | | | | N= 9041 R=0.7297 | N= 6189 R=0.8165 |
| SFX Clickthroughs | | | | | N= 8075 R=0.7295 |

Tables 4 through 9 show the correlations over the six journal value indicators in each of the six subject discipline journal subsets included in the study. These are biosciences (Table 4), social sciences (Table 5), engineering (Table 6), literature (Table 7), chemistry (Table 8), and history and philosophy (Table 9).

| Table 4: Bioscience Journal Value Indicator Correlations | | | | | |
|---|---|---|---|---|---|
| | UIUC Author Cited by outside authors | Articles written by UIUC authors | Downloads of articles by UIUC users | SFX Clickthroughs | Normalized Eigenfactor |
| Locally Cited by UIUC authors | N= 1204 R=0.8831 | N= 1204 R=0.5401 | N= 922 R=0.7760 | N=1164 R=0.5554 | N= 1162 R=0.7659 |
| UIUC Author Cited by outside authors | | N= 1204 R=0.5637 | N= 922 R=0.7031 | N= 1164 R=0.5978 | N= 1162 R=0.6591 |
| Articles written by UIUC authors | | | N= 922 R=0.4025 | N= 1164 R=0.5713 | N= 1162 R=0.3956 |
| Downloads of articles by UIUC users | | | | N= 909 R=0.6597 | N= 902 R=0.8420 |
| SFX Clickthroughs | | | | | N= 1128 R=0.5465 |

| Table 5: Social Science Journal Value Indicator Correlations | | | | | |
|---|---|---|---|---|---|
| | UIUC Author Cited by outside authors | Articles written by UIUC authors | Downloads of articles by UIUC users | SFX Clickthroughs | Normalized Eigenfactor |
| Locally Cited by UIUC authors | N= 1123 R=0.6199 | N= 1123 R=0.5059 | N= 938 R=0.5416 | N= 1100 R=0.6436 | N= 201 R=0.6130 |
| UIUC Author Cited by outside authors | | N= 1123 R=0.5156 | N= 938 R=0.5381 | N= 1100 R=0.5914 | N= 201 R=0.5209 |
| Articles written by UIUC authors | | | N= 938 R=0.4605 | N= 1100 R=0.5780 | N= 201 R=0.3595 |
| Downloads of articles by UIUC users | | | | N= 931 R=0.7231 | N= 173 R=0.6414 |
| SFX Clickthroughs | | | | | N= 196 R=0.7541 |

| Table 6: Engineering Journal Value Indicator Correlations | | | | | |
|---|---|---|---|---|---|
| | UIUC Author Cited by outside authors | Articles written by UIUC authors | Downloads of articles by UIUC users | SFX Clickthroughs | Normalized Eigenfactor |
| Locally Cited by UIUC authors | N= 1065 R=0.8533 | N=1065 R=0.8392 | N= 817 R=0.7240 | N= 1022 R=0.7046 | N= 1024 R=0.7401 |
| UIUC Author Cited by outside authors | | N= 1065 R=0.8021 | N= 817 R=0.7170 | N= 1022 R=0.6738 | N= 1024 R=0.8048 |
| Articles written by UIUC authors | | | N= 817 R=0.6002 | N= 1022 R=0.6288 | N= 1024 R=0.6272 |
| Downloads of articles by UIUC users | | | | N= 802 R=0.8249 | N= 791 R=0.7767 |
| SFX Clickthroughs | | | | | N= 985 R=0.7080 |

| Table 7: Literature Journal Value Indicator Correlations | | | | | |
|---|---|---|---|---|---|
| | UIUC Author Cited by outside authors | Articles written by UIUC authors | Downloads of articles by UIUC users | SFX Clickthroughs | Normalized Eigenfactor |
| Locally Cited by UIUC authors | N= 366 R=0.8846 | N= 366 R=0.4173 | N= 272 R=0.6859 | N= 354 R=0.5120 | N= 21 R=0.7530 |
| UIUC Author Cited by outside authors | | N= 366 R=0.4474 | N= 272 R=0.7652 | N= 354 R=0.6136 | N= 21 R=0.8057 |
| Articles written by UIUC authors | | | N= 272 R=0.5201 | N= 354 R=0.5204 | N= 21 R=0.7753 |
| Downloads of articles by UIUC users | | | | N= 268 R=0.7438 | N= 14 R=0.8971 |
| SFX Clickthroughs | | | | | N= 20 R=0.7659 |

| Table 8: Chemistry Journal Value Indicator Correlations | | | | | |
|---|---|---|---|---|---|
|  | UIUC Author Cited by outside authors | Articles written by UIUC authors | Downloads of articles by UIUC users | SFX Clickthroughs | Normalized Eigenfactor |
| Locally Cited by UIUC authors | N= 747 R=0.8557 | N= 747 R=0.9291 | N= 575 R=0.8667 | N= 722 R=0.7755 | N= 728 R=0.8207 |
| UIUC Author Cited by outside authors |  | N= 747 R=0.8960 | N= 575 R=0.8111 | N= 722 R=0.7922 | N= 728 R=0.8481 |
| Articles written by UIUC authors |  |  | N= 575 R=0.8871 | N= 722 R=0.8592 | N= 728 R=0.8540 |
| Downloads of articles by UIUC users |  |  |  | N= 570 R=0.9039 | N= 569 R=0.9130 |
| SFX Clickthroughs |  |  |  |  | N= 707 R=0.8370 |

| Table 9: History & Philosophy Journal Value Indicator Correlations | | | | | |
|---|---|---|---|---|---|
|  | UIUC Author Cited by outside authors | Articles written by UIUC authors | Downloads of articles by UIUC users | SFX Clickthroughs | Normalized Eigenfactor |
| Locally Cited by UIUC authors | N= 160 R=0.4363 | N= 160 R=0.5213 | N= 96 R=0.4107 | N= 155 R=0.5640 | N= 15 R=0.8030 |
| UIUC Author Cited by outside authors |  | N= 160 R=0.1516 | N= 96 R=0.1554 | N= 155 R=0.1762 | N= 15 R=0.7198 |
| Articles written by UIUC authors |  |  | N= 96 R=0.7735 | N= 155 R=0.8462 | N= 15 R=0.3774 |
| Downloads of articles by UIUC users |  |  |  | N= 96 R=0.8702 | N= 11 R=0.2085 |
| SFX Clickthroughs |  |  |  |  | N= 15 R=0.8331 |

Several of the correlation relationships, both in the all-discipline set and the subject subsets, bear further examination. The specific disciplinary correlations exhibit some interesting differences with the all-discipline analysis, particularly in three pairwise relationships: *downloads* and *locally cited*; *articles written* and *locally cited*; and *downloads* and *articles written*.

### Downloads **and** Locally Cited

Numerous studies have shown that the relationship between usage and citation is very complex, particularly at the article or paper level. [33] Issues involving usage and citation obsolescence characteristics, where the half-life of the articles that are downloaded and the median cited half-life are significantly different play a key role in the relationship between usage and citation. [34] At the article level, the articles appearing at the top of a citation ranking are not necessarily the most frequently downloaded articles, and vice versa. [35] Several researchers have noted that different disciplines have different citation practices and protocols concerning citation behavior. Citation habits differ from one scientific area to another and that there are several reasons for both citing an article and for downloading an article. [36] Importantly, it has been established that correlations between downloads and citations are higher when calculated at the journal title level than at the article level. [37]

This study uses journal title level metrics, with two combined years of download data and a combined five years of both local citation and publication authorship data. Within this more simplified approach, in the all-disciplines overarching set, the often-studied correlation between the indicator pair *downloads* and *locally cited* is highly significant at N=9190, R=0.7843. In the subject discipline analyses, the correlation is also high in the biosciences (N=922, R=0.7760), engineering (N=817, R=0.7240), chemistry (N=575, R=0.8667), and literature (N=272, R=0.6859), but lower in the social sciences subset (N=938, R=0.5416) and the history and philosophy (N=96, R=0.4107) journals subset. Even given the complexity, in this study using two years of download data and five years of local citations at the journal title level, the correlations were high overall and high in most of the six subject subsets.

Vaughan, Tang, and Yang analyzed 150 journals in 69 fields and found higher correlations between downloads and citations in the social sciences and humanities fields than in science, engineering, and medicine fields. [38] In this study, the sciences and engineering fields yielded the highest correlations and the social sciences and humanities (except for literature) were lower.

Within the scholarly communications system, researchers are, for the most part, citing the most relevant and important articles in their field and faculty and students are downloading the most relevant articles for their research and instruction. Moed and Halevi suggested that for downloads and citations, there was a high correlation between the two in specialized fields in which the readers tend to be the active researchers but in fields where the reader population is wider and more diverse than the research community, the correlations are lower. [39] That observation is generally supported by the results of this analysis. It is also the case that some less academic articles in more general journals are being downloaded for classroom use and not research.

### Articles Written and Locally Cited

While the overarching all-disciplines correlation between the values in the *articles written and locally cited* pair is significantly high at N=12200 and R=0.7698, there are clear differences in the values derived in the six subject discipline subsets. Looking at Tables 4 through 9, the biosciences journals (N=1204, R=0.5401) in Table 4, the social sciences journals (N=1123, R=0.5059) in Table 5, the literature journal subset (N=366, R=0.4173) in Table 7, and the history and philosophy journals (N=160, R=0.5213) in Table 9 are all below the all-disciplines value. The engineering publications subset shown in Table 6 (N=1065, R=0.8392) and chemistry in Table 8 (N=747, R=0.9291) exhibit higher correlations than the other disciplines.

From the scholarly communications standpoint, the faculty are citing the most important articles in the most prestigious journals in the bibliographies of their research and, at the same time, are trying to have their research published in those most prestigious journals. Research faculty aspire to be published in the same journals that are publishing the most highly cited articles. The all-discipline correlation between the indicators *articles written* and *locally cited* is very high at R=0.7698 and that shows a significant university-wide ability to publish in the same journals that are being cited. But the correlation varies across the six subsets. Interestingly, the *articles written* and *locally cited* correlations in the six subject discipline subsets examined in this study match very closely with the associated program rankings in the U.S. News and World Report 2022 Graduate School Rankings where the UIUC chemistry program is ranked at #6, engineering is #10 (with computer science at #5), biological sciences is #27, sociology is #49 and social work is #22 (but psychology is #7), English is #20 with literary criticism at #18, and history is ranked #21. In fact, one could argue that perhaps a factor in the determination of the prestige of a department or program could be the strength of the correlation between a faculty group's *articles written* and their *locally cited* indicator pair. This might be incorporated into the suite of algorithms of the ranking services. Looking at the six subset areas in this case, UIUC researchers in some of the higher ranked science and engineering programs look to be better able to publish in the journals that they are citing most frequently.

### *Downloads and Articles Written*

From this study, it appears that downloads are not strongly predictive of local authorship, given the all-disciplines correlation of the indicator pair *downloads* and *articles written* at N=9190, R=0.5282. In the subject discipline subsets, there are three higher correlation values in the history and philosophy journals (N=96, R=0.7735), chemistry journals (N=575, R=0.8871), and engineering journals (N=817, R=0.6002). The other three subject discipline analyses show lower correlations with biosciences (N=922, R=0.4205) and social science (N=938, R=0.4605) lower, and literature (N=272, R=0.5201) somewhat lower. An examination of the five correlations involving the downloads values shows that the correlations for the pair *downloads* and *articles written* are typically lower in the all-discipline journals and the subset disciplines than the other correlations involving downloads. These lower relationships may be due to the same issue contributing to the other lower correlations involving the articles written measures. They are related to the aspirational aspects of UIUC authorship, where the researchers may not be able to consistently publish, because of low acceptance rates or quality of their research, in the journals that, in this case, contain those articles that faculty and students are downloading to support their research. It is also possible that in the broader or more popular fields, there may be numerous downloads of articles by non-researchers in the field. However, in that case the history and philosophy correlation values would be expected to be low as they were for *downloads* and *locally cited* pair but here, they are in fact the highest in the disciplinary set.

Overall, the highest download numbers and local citation numbers come from many of the prestigious journals where the faculty aspire to publish.

*SFX Local Link Resolver*

De Groote, Blecic, and Martin defined the term SFTARs (successful full-text article requests) to indicate how many times articles in a journal are retrieved from a local link resolver full-text link appearing in an abstracting and indexing (A&I) service. [40] Their study of medical journals found significant correlations between local link resolver requests and local citations. In this study, the correlation between *SFX clickthroughs* and *locally cited* pair in the all-discipline journal set was moderate (N=11709, R=0.5863). Interestingly, the *SFX clickthroughs* and *locally cited* pair values are highest in the social science journals (N=1100, R=0.6436), engineering journals (N=1022, R=0.7046), and chemistry journals (N=722, R=0.7755).

The two highest correlations involving the SFX local link resolver clickthroughs in the all-discipline journals are *SFX clickthroughs* and *downloads* (N=9041, R=0.7297) and *SFX clickthroughs* and *normalized Eigenfactor* (N=8075, R=0.7295). The other values, including the correlations with articles written and local citations were lower.

In the UIUC environment, the use of local link resolver links is reduced by several factors. Some users go directly to a publisher or journal site and bypass any link resolver usage. In addition, some full-text links appear on aggregator sites and in discovery systems offering users direct to full-text links or direct to publisher site links. Direct publisher site links appear in major A&I services such as Scopus, PubMed, or Web of Science and aggregators and discovery systems offer direct full-text links or DOI links to publisher sites, both of which bypass the local link resolver.

It may be that UIUC SFX usage was not uniform across all A&I services and publisher sites and that it was consistently used only in certain subject A&I services and not in others or that users were clicking on the direct to PDF links in some A&I services rather than the SFX links. This has gotten more complicated in the UIUC Library where the discovery service pulls out direct PDF links from EBSCOHOST and ProQuest services. In the current environment, the SFX local link resolver has been replaced by the Alma link resolver.

*External Citation Values*

The LJUR data provides external citation values for the journal title articles authored by UIUC researchers that are cited by outside researchers. In the all-discipline set of journals, the correlation for the pair *outside citations* and *locally cited* (N=12200, R=0.7613) is high and it is significantly high for all the disciplinary subsets except for history and philosophy (N=160, R=0.4363).

The correlations for the *outside citations* and *downloads* pair (N=9190, R=0.4959) were low in the all-disciplines set but higher in the biosciences, engineering, literature, and chemistry subsets. The correlation for *outside citations* and *SFX clickthroughs* (N=11709, R=0.4351) was lower in the all-discipline journals than it was in all the subset journal collections except for history and philosophy.

The correlation value for the *outside citations* and *articles written* pair (N=12200, R=0.7907) in the overarching all-discipline collection is significantly high but the *outside citations* and *articles written* pair exhibits the exact same differences in correlation values within the six subject subsets that were present in the *local citations* and *articles written* values.

Overall, the external citation correlations do not appear to contribute to a better understanding of the relationships between publication, citation, and usage metrics. The fact that the *outside citations* and *articles written* pair exhibited the same subject subset differences as the *local citations* and *articles written* pair in the six journal sets again implies that UIUC faculty in some departments or programs are not always writing in the same highly regarded journals that they are citing or that outside researchers are citing.

### Normalized Eigenfactor

The normalized Eigenfactor Score values are the only global impact factor measures that exhibit significant correlation values with local publication, citation, and download data. This is particularly true at the all-disciplines level. The *normalized Eigenfactor* and *locally cited* pair (N=8408, R=0.7858) and *normalized Eigenfactor* and *downloads* pair (N=6189, R=0.8165) are the highest all-discipline correlations. Several of the *Eigenfactor* and *articles written* correlations in the subject subsets are low, with the biosciences (N=1162, R=0.3956), social sciences (N=201, R=0.3595), and history and philosophy (N=15, R=0.3774) subsets exhibiting low pairwise correlation values.

## Limitations

The study used publication and citation data from 2013 to 2017 and download data from 2015 and 2018 in order to accommodate the projected half-life and obsolescence issues connected with the complex relationships between usage, citations, and publications. This placed the study in a time period before open access became as prevalent as it is currently. The implications of open access full-text downloads and authoring are not known but should be investigated using later years for the study.

The authorship, citation and download data numbers are all raw numbers and are not log normalized or weighted. There is no weight given to first or last authors listed on the articles and all cited articles are treated the same. It is not clear if weighting would influence the correlations in any way.

The study looked at only six subject subset areas. There is a clear need to examine the metric correlations within additional disciplines – some of the other locally highly ranked subject areas and some of the lesser ranked programs – in order to see if the conclusions regarding program strength and the relationship between the *articles written* and *locally cited* parameters and several other pairs hold true. It would be possible to automate the process to introduce a script that would present the appropriate SQL commands to derive the subject discipline subsets to calculate the R values and summarize and collect the results.

The COUNTER only publisher full-text download data encompassed 31,918 journal titles but the ISI LJUR coverage extended to 17,934 total journals including 12,200 active journals. The Scopus API journal coverage includes almost 25,000 current journals and would be more extensive than the coverage provided by the LJUR data. Repeating this study using UIUC authored journal articles and processed using the Scopus API would provide more extensive journal title coverage and allow additional journal metric pair correlations to be performed.

## Conclusions

The goal of this study was to investigate the correlational relationships between journal title metrics from the UIUC multi-disciplinary research journal collection and over six subject subset journals in biosciences; chemistry; social sciences; history and philosophy; literature; and engineering. The particular metric indicators making up this analysis were local publication and citation data; COUNTER supplied full-text downloads; local link resolver clickthroughs; and four global impact factor index values. This analysis was carried out over a large sample of 12,200 active journals in all subject disciplines with publication and citation data supplied through the ISI LJUR service. Full-text download numbers from COUNTER were available for 9,190 journal titles in the active journal title set.

The exercise of assembling the raw journal title publication, citation, and download values over a collection-scale set of journals was useful in itself. A web interface over the database table was created with a search function that allowed retrieval by journal title, publisher, and subject with sorting capabilities by each of the journal title metrics. The web site tool provides data on individual journal titles which can inform subscription, retention, and cancellation decisions, assist liaison librarians in understanding department and group research concentrations, and could contribute to the generation of core journal lists. The pairwise correlation values over the journal title metrics provide insight into scholarly communication patterns, the relationships between the various journal metrics, and the bibliometric interactions in operation at the UIUC Library. These correlation values can be compared to the values at other R1 university libraries. They can also provide evidence of the ability of one or more of the metrics to be used as a proxy for the others. The process methodology and protocols for this study can serve as a model or blueprint for other academic libraries looking to investigate these relationships in other institutional settings.

An analysis of the four global impact factor indices showed that the ISI JCR, the SNIP, and the SCIMago JCR indices did not exhibit significant positive correlations with the publication, citation, or download indicators. Only the normalized Eigenfactor values showed significant correlation with the local data.

The relationship between local download usage and local citation has been the subject of many previous investigations. Earlier studies have shown that the relationships involving downloads and citations, particularly when they are examined at the article level rather than the journal title level, are quite complex. The data in this study were comprised of two combined years of download data from 2015 and 2018 and a combined five years (from 2013 to 2017) of both local and external citation and publication authorship data. The correlations between the important *downloads* and *locally cited* values were calculated at the journal title level, where it has been shown to be higher than at the article level. The analysis found an overall strong positive correlation between journal usage, in the form of full-text downloads, and locally cited journal titles. In the all-disciplines overarching journal set, the correlation between *downloads* and *locally cited* pair was high (N=9190, R=0.7843) and the R values were also high (from 0.5416 to 0.8667) in five of the six subject subset journal collections examined in the study. The history and philosophy subset R value was 0.4107. While earlier investigations have proven inconclusive, this study shows strong correlations in the all-disciplines set and most subject subsets between full-text downloads and local citations.

One explanation offered in the literature for the subject discipline differences may lie in the observation that there are higher correlations between the two metrics in specialized fields in which the readers tend to be the active researchers but lower correlations in fields where the reader population is wider and more diverse than the research community. Within the all-disciplines 12,200 active journals, and in most subject disciplines, this study's correlation results do imply that download measurements can predict local citations and vice versa.

Researchers are citing the most important articles in the most prestigious journals in their field. At the same time, they are attempting to publish their research in the most prestigious journals, which are typically the journals that they and other researchers are predominantly citing. Research faculty aspire to be published in the same journals that are publishing the most highly cited articles. The all-discipline correlation of the *articles written* and *locally cited* is very high at R=0.7698, demonstrating a significant ability of UIUC faculty to publish in the same journals that are citing. This study found, however, that, for the six subject journal title subsets that are identified from the overarching collection, the *articles written* and *locally cited* correlation for these six subject discipline subsets matches very closely with their associated department or program ranking in the U.S. News and World Report 2022 Graduate Schools Rankings. One criterion for ranking a subject department or program might be to calculate the strength of the correlation between the group's *articles written* and *locally cited* journal title metrics. The aspirational publishing aspect may also affect the correlations between the indicators *downloads* and *articles written* where many of the articles are downloaded from highly cited journals where researchers aspire to publish.

The study found that SFX link resolver correlations were high when matched with the *downloads* indicator and the *normalized Eigenfactor* measures. The link resolver correlation values with articles written and local citations were lower. The link resolver and external citation indicators were not regarded as very useful measures for understanding publication, citation, and usage behaviors or activities.

With the addition of journal title subscription information to the metric data assembled in this study, it is fairly easy to calculate a journal composite value, using the weighted set of local publication, citation, and download number values to derive a journal composite value which can then be divided by the subscription price to obtain an overall value score. The UIUC Library has produced this assigned value journal table, although there is some difficulty in assigning an individual journal subscription price to journals purchased as part of an overarching "big deal" package.

The study revealed some interesting interactions and relationships between the journal metrics. There are limitations and subtleties with each of the journal title measure correlations. Chew et al noted that "it is generally conceded that the metrics, when taken in aggregate, provide a more complete picture on journal value and importance." [41] A number of studies show that the various journal metrics need to be applied and combined in a strategic manner in order to obtain meaningful results [42] De Groote, Blecic, and Martin noted that citation data describes research activity but that vendor, publisher, and link-resolver statistics also reflect educational and clinical usage. [43] Given these complex and interrelated factors and the analysis presented in these study results, it would appear that multiple metrics may need to be employed to make definitive statements about journal publication, citation, and usage relationships and interactions. It is also clear from the study that a more nuanced profile of user publication, citation, and usage activity than some other measures, such as the commonly quoted cost per use metric, are possible and desirable.

## Notes

1. Davis, Philip M., "Where to Spend our E-journal Money? Defining a University Library's Core Collection Through Citation Analysis". *Portal: Libraries & the Academy* vol: 2, issue 1, 2002, pp. 155-166. doi:10.1353/pla.2002.0009.

2. Haddow, Gaby, "Level 1 COUNTER Compliant Vendor Statistics are a Reliable Measure of Journal Usage" *Evidence Based Library & Information Practice* vol: 2, issue 2, 2007, pp. 84-86 https://doi.org/10.18438/B83G6S

3.  Hoffmann, Kristin and Doucette, Lise, "A Review of Citation Analysis Methodologies for Collection Management". *College & Research Libraries* vol: 73, issue 4, 2012, pp. 321-335. https://doi.org/10.5860/crl-254

4.  Ashman, Allen B., "An Examination of the Research Objectives of Recent Citation Analysis Studies". *Collection Management* vol: 34, issue 2, 2009, pp. 112-128. https://doi.org/10.1080/01462670902725885

5.  Hoffmann and Doucette; "A Review of Citation Analysis Methodologies for Collection Management." 327.

6.  McGillivray, Barbara.; Astell, Mathias., "The relationship between usage and citations in an open access mega-journal". *Scientometrics* vol: 121, issue 2, 2019, pp. 818. https://doi.org/10.1007/s11192-019-03228-3

7.  Pastva, Joelen; Shank, Jonathan; Gutzman, Karen E.; Kaul, Madhuri; Kubilius, Ramune K., "Capturing and Analyzing Publication, Citation, and Usage Data for Contextual Collection Development". *Serials Librarian* vol: 74, issue 1-4, 2018, p. 102-110. https://doi.org/10.1080/0361526X.2018.1427996

8.  Brody, Tim; Harnad, Stevan; Carr, Leslie, "Earlier Web usage statistics as predictors of later citation impact". *Journal of the American Society for Information Science & Technology* vol: 57, issue 8, 2006, pp. 1060-1072. https://doi.org/10.1002/asi.20373

9.  Kurtz, Michael J.; Eichhorn, Guenther; Accomazzi, Alberto; Grant, Carolyn; Demleitner, Markus; Murray, Stephen S.; Martimbeau, Nathalie; Elweli, Barbara. "The bibliometric properties of article readership information." *Journal of the American Society for Information Science and Technology*, 56, 2005, pp. 111–128. https://doi.org/10.1002/asi.20096

10. Kurtz, Michael J.; Bollen, Johan, "Usage Bibliometrics". *Annual Review of Information Science and Technology*, vol: 44, 2010, p. 1-64. https://doi.org/10.1002/aris.2010.1440440108

11. Ibid. 23.

12. Schloegl, Christian.; Gorraiz, Juan., "Comparison of citation and usage indicators: The case of oncology journals." *Scientometrics*. vol. 82 Issue 3, 2010, pp. 567-580 https://doi.org/10.1007/s11192-010-0172-1; Schloegl, Christian.; Gorraiz, Juan, "Global usage versus global citation metrics: The case of pharmacology journals." *Journal of the American Society for Information Science & Technology* vol: 62, issue 1, 2011, pp. 161-170. https://doi.org/10.1002/asi.21420

13. Schloegl and Gorraiz, "Comparison of citation and usage indicators: The case of oncology journals."; Gorraiz, Juan; Gumpenberger, Christian; Schlogl, Christian, "Usage versus citation behaviours in four subject areas." *Scientometrics*. vol. 101, no. 2 (November 2014), p. 1077-1095. https://doi.org/10.1007/s11192-014-1271-1 ; Vaughan, Liwen; Tang, Juan.; Yang, Rongbin., "Investigating disciplinary differences in the relationships between citations and downloads.", *Scientometrics*. 1 June 2017, Vol. 111 Issue 3, 1533-1545. https://doi.org/10.1007/s11192-017-2308-z ; Wical, Stephanie H.; Vandenbark, R. Todd, "Combining Citation Studies and Usage Statistics to Build a Stronger Collection." *Library Resources & Technical Services* vol: 59, issue 1, 2015, 33-42. https://doi.org/10.5860/lrts.59n1.33; Bollen Johan.; Van De Sompel, Herbert;

Smith Joan A., Luce Rick., "Toward alternative metrics of journal impact: A comparison of download and citation data.", *Information Processing and Management*. December 2005, Vol. 41 Issue 6, 1419-1440. https://doi.org/10.1016/j.ipm.2005.03.024; Chu, H. & Krichel, T., "Downloads vs. citations in economics: Relationships, contributing factors and beyond *Proceedings of ISSI 2007 - 11th International Conference of the International Society for Scientometrics and Informetrics, (2007), 207-215* https://www.issi-society.org/publications/issi-conference-proceedings/proceedings-of-issi-2007/ ; O'Leary "The relationship between citations and number of downloads in Decision Support Systems", *Decision Support Systems, (2008), 45(4), 972-980.* https://doi.org/10.1016/j.dss.2008.03.008 ; Dewland, Jason C., "A Local Citation Analysis of a Business School Faculty: A Comparison of the Who, What, Where, and When of Their Citations", *Journal of Business & Finance Librarianship* vol: 16, issue 2, 2011, 145-158. https://doi-org/10.1080/08963568.2011.554740 : Ke, Irene; Bronicki, Jackie, "Using Scopus to Study Researchers' Citing Behavior for Local Collection Decisions: A Focus on Psychology." *Journal of Library Administration* vol: 55, issue 3, 2015, 165-178. https://doi-org/10.1080/01930826.2015.1034035

14. Moed, Henk F.; Halevi, Gali, "On full text download and citation distributions in scientific-scholarly journals." *Journal of the Association for Information Science & Technology* vol: 67, issue 2, 2016, 412-431. https://doi-org.proxy2.library.illinois.edu/10.1002/asi.23405; Chew, Katherine, Schoenborn, Mary, Stemper, James., Lilyard, Caroline., "E-journal metrics for collection management: Exploring disciplinary usage differences in Scopus and Web of Science." Evidence Based Library and Information Practice. Vol. 11 Issue 2, 2016, 97-120. https://doi.org/10.18438/B85P87 ; Guerrero-Bote, Vincente P.; Moya-Anegón, Felix., "Relationship between downloads and citations at journal and paper levels, and the influence of language" Scientometrics vol: 101, issue 2, 2014, 1043-1065. https://doi.org/10.1007/s11192-014-1243-5

15. Duy, Joanna; Vaughan, Liwen, "Can Electronic Journal Usage Data Replace Citation Data as a Measure of Journal Use? An Empirical Examination" Journal of Academic Librarianship vol: 32, issue 5, 2006, 512-517. https://doi.org/10.1016/j.acalib.2006.05.005

16. McDonald, John D., "Understanding journal usage: A statistical analysis of citation and use." Journal of the American Society for Information Science & Technology vol: 58, issue 1, 2007, 39-50. https://doi.org/10.1002/asi.20420

17. De Groote, Sandra L.; Blecic, Deborah D.; Martin, Kristin E., "Measures of health sciences journal use: a comparison of vendor, link-resolver, and local citation statistics." Journal of the Medical Library Association vol: 101, issue 2, 2013, 110-119. https://doi.org/10.3163/1536-5050.101.2.006

18. Gallagher John.; Bauer Kathleen.; Dollar Daniel M., "Evidence-based librarianship: Utilizing data from all available sources to make judicious print cancellation decisions." Library Collections, Acquisition and Technical Services. June 2005, Vol. 29 Issue 2, 169-179. https://doi.org/10.1016/j.lcats.2005.04.004

19. Chew et al, "E-journal metrics for collection management: Exploring disciplinary usage differences in Scopus and Web of Science."

20. Pastva et al, "Capturing and Analyzing Publication, Citation, and Usage Data for Contextual Collection Development.", 102-110

21. Gao, Wenli, "Beyond journal impact and usage statistics: Using citation analysis for collection development.", Serials Librarian vol: 70, issue 1-4, 2016, 121-127. https://doi.org/10.1080/0361526X.2016.1144161

22. Bollen, Johan; Van de Sompel, Herbert; Hagberg, Aric; Chute, Ryan, "A Principal Component Analysis of 39 Scientific Impact Measures." PLoS ONE vol: 4, issue 6, 2009, 1-11. https://doi.org/10.1371/journal.pone.0006022

23. Ibid. 10.

24. Bollen et al, "Toward alternative metrics of journal impact: A comparison of download and citation data."

25. Gorraiz, Gumpenberger, and Schlogl, "Usage versus citation behaviours in four subject areas."

26. Ibid., 1093.

27. Mark R. Elkins, Christopher G. Maher, Robert D. Herbert, Anne M. Moseley and Catherine Sherrington, "Correlation between the Journal Impact Factor and three other journal citation indices", *Scientometrics*. 2010, Vol. 85 Issue 1, 81-93. https://doi.org/10.1007/s11192-009-0090-2

28. Moed and Halevi "On full text download and citation distributions in scientific-scholarly journals."

29. Knowlton, Steven A.; Sales, Adam C.; Merriman, Kevin W., "A Comparison of Faculty and Bibliometric Valuation of Serials Subscriptions at an Academic Research Library." *Serials Review* vol: 40, issue 1, 2014, 28-39. https://doi-org/10.1080/00987913.2014.897174

30. Jurczyk, Eva; Jacobs, Pamela, "What's the Big Deal? Collection Evaluation at the National Level." *Portal: Libraries & the Academy* vol: 14, issue 4, 2014, 617-631. https://doi.org/10.1353/pla.2014.0029

31. Bollen et al, "A Principal Component Analysis of 39 Scientific Impact Measures."; Bollen et al, "Toward alternative metrics of journal impact: A comparison of download and citation data."; Chew et al, "E-journal metrics for collection management: Exploring disciplinary usage differences in Scopus and Web of Science."; Schloegl and Gorraiz, "Comparison of citation and usage indicators: The case of oncology journals."; Schloegl and Gorraiz "Global usage versus global citation metrics: The case of pharmacology journals."; Altmann Klaus G.; Gorman G. E., "Can impact factors substitute for the results of local use studies? Findings from an Australian case study", *Collection Building*. 1 June 1999, Vol. 18 Issue 2, 90-94. https://doi-org/10.1108/01604959910265878 Davis, "Where to Spend our E-journal Money? Defining a University Library's Core Collection Through Citation Analysis"; Kreider, "The correlation of local citation data with citation data from 'Journal Citation Reports'" *Library Resources & Technical Services* vol: 43, issue 2, 1999, 67-77. https://doi.org/10.5860/lrts.43n2.6 ; Stephen P. Harter and Thomas E. Nisonger, "ISI's Impact Factor as Misnomer: A Proposed New Measure to Assess Journal Impact Factor," *Journal of the American Society for Information Science* 48 (Dec. 1997): 1146–48. https://doi.org/10.1002/(SICI)1097-4571(199712)48:12<1146::AID-ASI9>3.0.CO;2-U

32. Chew et al, "E-journal metrics for collection management: Exploring disciplinary usage differences in Scopus and Web of Science."

33. Kurtz and Bollen, "Usage Bibliometrics"; Schloegl and Gorraiz, "Comparison of citation and usage indicators: The case of oncology journals"; Schloegl and Gorraiz "Global usage versus global citation metrics: The case of pharmacology journals"; Moed and Halevi "On full text download and citation distributions in scientific-scholarly journals"; Gorraiz, Gumpenberger, and Schlogl, "Usage versus citation behaviours in four subject areas."

34. Kurtz et al "The bibliometric properties of article readership information"; Kurtz and Bollen, "Usage Bibliometrics"; Schloegl and Gorraiz, "Comparison of citation and usage indicators: The case of oncology journals"; Schloegl and Gorraiz "Global usage versus global citation metrics: The case of pharmacology journals."

35. Moed and Halevi "On full text download and citation distributions in scientific-scholarly journals"; Gorraiz, Gumpenberger, and Schlogl, "Usage versus citation behaviours in four subject areas."

36. Guerrero-Bote and Moya-Anegón, "Relationship between downloads and citations at journal and paper levels, and the influence of language"; Nederhof, A. J., Zwaan, R. A., De Bruin, R. E., & Dekker, P. J., Assessing the usefulness of bibliometric indicators for the humanities and the social sciences. *Scientometrics,* 15, 1989, 423–435. https://doi.org/10.1007/BF02017063 ; Gorraiz, Gumpenberger, and Schlogl, "Usage versus citation behaviours in four subject areas"; Kurtz and Bollen, "Usage Bibliometrics."

37. Moed and Halevi "On full text download and citation distributions in scientific-scholarly journals"; Vaughan, Tang, and Yang, "Investigating disciplinary differences in the relationships between citations and downloads"; Moed, Henk F., "Statistical relationships between downloads and citations at the level of individual documents within a single journal." *Journal of the American Society for Information Science & Technology* vol: 56, issue 10, 2005, 1088-1097. https://doi-org/10.1002/asi.20200 ; Guerrero-Bote and Moya-Anegón, "Relationship between downloads and citations at journal and paper levels, and the influence of language."

38. Vaughan, Tang, and Yang, "Investigating disciplinary differences in the relationships between citations and downloads."

39. Moed and Halevi "On full text download and citation distributions in scientific-scholarly journals."

40. De Groote, Blecic, and Martin, "Measures of health sciences journal use: a comparison of vendor, link-resolver, and local citation statistics."

41. Chew et al, "E-journal metrics for collection management: Exploring disciplinary usage differences in Scopus and Web of Science," p.99.

42. Chew et al, "E-journal metrics for collection management: Exploring disciplinary usage differences in Scopus and Web of Science"; Wical and Vandenbark, "Combining Citation Studies and Usage Statistics to Build a Stronger Collection."

43. De Groote, Blecic, and Martin, "Measures of health sciences journal use: a comparison of vendor, link-resolver, and local citation statistics."